\documentclass[letterpaper,pra,floatfix,twocolumn,aps,10pt]{revtex4-1}

\usepackage{
	amssymb, 
	amsthm, 
	graphicx, 
	xspace,
	overpic,
	ifthen,
	verbatim, 
	subfigure, 
	appendix,
	ifthen,
	nicefrac, 
	color, 
	tensor, 
	etoolbox, 
	centernot, 
	MnSymbol, 
	amsmath,
	bm
} 

\newcommand{\bra}[1]{\langle #1 \rvert  }
\newcommand{\ket}[1]{\lvert #1 \rangle  }

\newcommand{\braket}[2]{  \langle #1 \vert #2 \rangle  }

\newcommand{\refstyle}{2}  
\newcommand{\eqrefstyle}{2}  
\newcommand{\Jeqref}[1] {\ifnum\refstyle=1{Eq.~(\ref{#1})}\else{equation (\ref{#1})}\fi}
\newcommand{\Jeqsref}[1] {\ifnum\refstyle=1{Eqs.~(\ref{#1})}\else{equations (\ref{#1})}\fi}
\newcommand{\EQref}[1] {\ifnum\eqrefstyle=1{(\ref{#1})}\else{\Jeqref{#1}}\fi}
\newcommand{\EQsref}[1] {\ifnum\eqrefstyle=1{(\ref{#1})}\else{\Jeqsref{#1}}\fi}
\newcommand{\Eqref}[1]  {\ifnum\eqrefstyle=3{\Jeqref{#1}}\else{(\ref{#1})}\fi}
\newcommand{\Eqsref}[1]  {\ifnum\eqrefstyle=3{\Jeqsref{#1}}\else{(\ref{#1})}\fi}
\newcommand{\Figref}[1]{\ifnum\refstyle=1{Fig.~\ref{#1}}\else{figure~\ref{#1}}\fi}
\newcommand{\Figsref}[1]{\ifnum\refstyle=1{Figs.~\ref{#1}}\else{figures~\ref{#1}}\fi}
\newcommand{\Appref}[1]{\ifnum\refstyle=1{App.~\ref{#1}}\else{appendix~\ref{#1}}\fi}
\newcommand{\Appsref}[1]{\ifnum\refstyle=1{Apps.~\ref{#1}}\else{appendices~\ref{#1}}\fi}
\newcommand{\Chref}[1]{\ifnum\refstyle=1{Ch.~\ref{#1}}\else{chapter~\ref{#1}}\fi}
\newcommand{\Chsref}[1]{\ifnum\refstyle=1{Chs.~\ref{#1}}\else{chapters~\ref{#1}}\fi}
\newcommand{\Secref}[1]{\ifnum\refstyle=1{Sec.~\ref{#1}}\else{section~\ref{#1}}\fi}
\newcommand{\Secsref}[1]{\ifnum\refstyle=1{Secs.~\ref{#1}}\else{sections~\ref{#1}}\fi}
\newcommand{\Refref}[1]{\ifnum\refstyle=1{Ref.~\cite{#1}}\else{reference~\cite{#1}}\fi}
\newcommand{\Refsref}[1]{\ifnum\refstyle=1{Refs.~\cite{#1}}\else{references~\cite{#1}}\fi}

\newcommand{\draftmode}{1}    
\newcommand{\notetoself}[1]{\ifnum \draftmode=1 {\color[rgb]{0,0,0.8} [#1]} \fi}  
\newcommand{\cuttext}[1]{\ifnum \draftmode=1 {\color[rgb]{0,0.5,0} [#1]} \fi}  
\newcommand{\warntext}[1]{\ifnum \draftmode=1 {\color[rgb]{0.9,0.6,0} #1} \else {#1} \color{black} \fi}  
\pdfoutput=1

\renewcommand{\draftmode}{0} 

\begin{document}



\title{Evidence for gravitons from decoherence by bremsstrahlung}
\date{\today}
\author{C.~Jess~Riedel}
\affiliation{IBM Watson Research Center, Yorktown Heights, NY 10598}


\begin{abstract}
It is thought that gravitons are impossible to detect, even with the technological ability to construct experiments larger than Jupiter.  However, in principle it is possible to detect the emission of single gravitons through the decoherence of relativistic Planck-mass superpositions by gravitational bremsstrahlung.  Although enormous experimental challenges ensure that such an experiment will not be achievable in the foreseeable future, this possibility suggests that gravitons are not forever outside our empirical grasp.  It is also evidence that decoherence as a detection method has untapped potential.
\end{abstract} 

\maketitle


Gravitons, the particle associated with the quantization of the gravitational force, have never been seen but are generally believed to exist.  Unfortunately, due to their very weak interactions, they are probably undetectable by any feasible measurement using traditional techniques \cite{DysonReview}.

Rothman and Boughn \cite{Rothman2006,Boughn2006} surveyed several possible sources of gravitons and considered hypothetical detectors built with advanced technology.  They showed that the highest luminosity in the galaxy would be found by parking a detector in close orbit around a neutron star.  Such a detector could not be much larger than Jupiter (for it to be supported electrostatically from gravitational collapse under its own weight) nor much closer than the Roche radius (lest it be torn apart by tidal forces).  Even at these fantastics limits, one would expect it to absorb no more than one graviton every decade.  Furthermore, it would be unlikely to be able to distinguish this from background events in any imaginable manner \cite{Rothman2006}.

In this letter, I suggest an alternate strategy based on detecting the graviton through the quantum decoherence it causes rather than any classical effect such as energy or momentum transfer.  This is a special case of the general idea of detecting weak (or rather \emph{soft}) phenomena through decoherence; it has been shown elsewhere to be sensitive to effects that are \emph{classically undetectable} (in a sense that can be made precise) \cite{Riedel02, Riedel03}.

A graviton decoherence experiment lies far beyond the reach of technology today or in the forseeable future.  The scope of this letter is only to show that it is possible in principle, and I will not catalog all of the massive difficulties that must be surmounted to build it.  My purpose is just to (a) illustrate the risk of prematurely concluding that certain feats are impossible because of failure of imagination and (b) gather evidence that decoherence detection is a promising and unexplored concept for the measurement of soft phenomena.





\begin{figure} [b!]
    \centering 
  \newcommand{\pbwidthfactor}{0.95} \includegraphics[width=\pbwidthfactor\columnwidth]{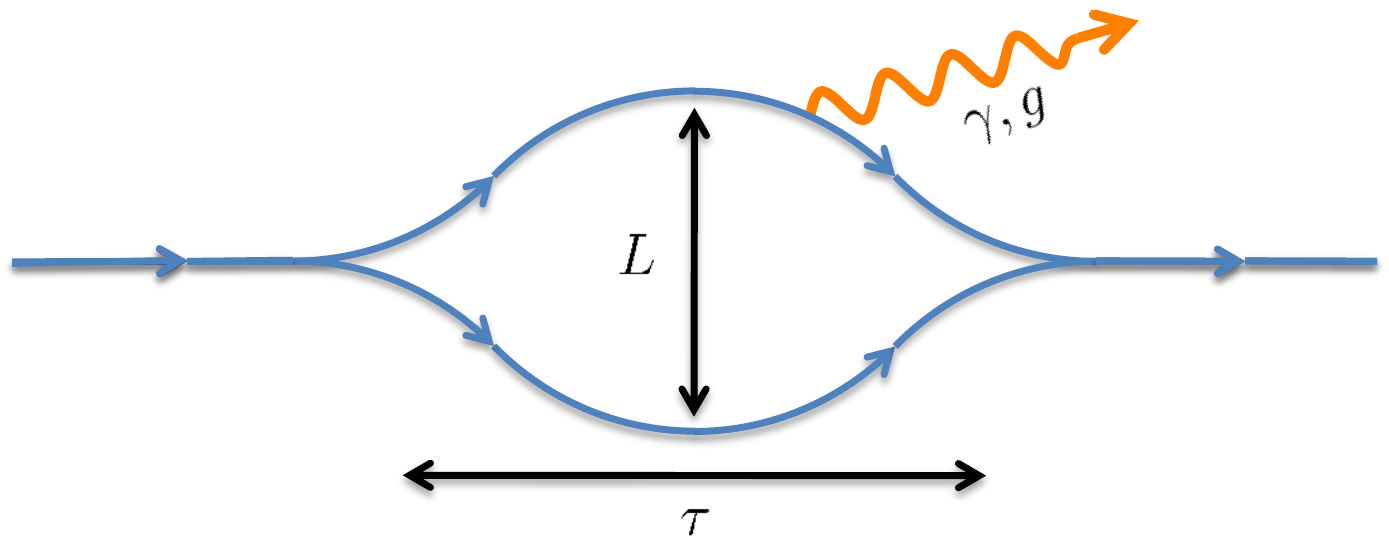}
  \caption{Decoherence by electromagnetic or gravitational bremsstrahlung. A charged object is brought into a coherent superposition of spatial extent $L$ then recombined after a time $\tau$.  If the two paths include sufficient relative acceleration for the given charge, they decohere through the emission of bremsstrahlung, which records with-path information.  The elementary charges or masses act together coherently to generate the radiation, which need not be absorbed for its existence to be detected.}
  \label{fig:wiggler_diagram}
\end{figure}

Consider the toy matter interferometer in figure \ref{fig:wiggler_diagram} in which a object with mass $m$ is brought into a superposition of two spatially localized center-of-mass wavepackets separated by a distance $L$ and then smoothly recombined after a time $\tau$, so that the interference pattern can be observed.  The two wavepackets achieve a typical relative speed of $v = \beta c \sim L/\tau$ (and acceleration $a \sim L/\tau^2$).  Even if there is no decoherence of the two paths from external environments, there can be substantial \emph{intrinsic} decoherence due to emitted radiation \cite{Breuer2000, Breuer2001, BreuerBrem, Durr2000}.  When the clump is thermal, there are two types of sources: blackbody radiation and bremsstrahlung. (I ignore emissions due to excited states, such as radioactive decays.)

Decoherence from blackbody radiation can be avoided by cooling the body to a temperature with characteristic wavelength much longer than $L$. Electromagnetic bremsstrahlung can be avoided by ensuring that there is no net electromagnetic charge.  (Microscopic charge inhomogeneities are smoothed out on scales smaller than $\lambda$.)  But the gravitational charge (i.e.\ mass) is always positive, so gravitational bremsstrahlung is irreducible for paths with a given acceleration.

Let us first consider the general process of decoherence by bremsstrahlung radiation of a massless field, and then specialize to electromagnetism and gravity in turn.  Importantly, we may treat the two paths taken by the wavepackets as classical sources so long as the following two conditions hold \cite{Breuer2001, ford1997electromagnetic, ford1993electromagnetic}.  (1) The energy of the emitted graviton is much less than rest-energy of the particles composing the object, i.e.\ that $\lambda \gg \lambda_{\mathrm{C}}$, where $\lambda_{\mathrm{C}}$ is the Compton wavelength of the electron.  This ensures that the energy radiated is very small compared to the kinetic energy so that the effects of radiative damping and pair creation/annihilation are negligible.  (2) The objects has a tightly defined velocity, i.e.\ that $\sigma_p/m \ll v$, where $\sigma_p$ is the spread of the wavefunction in momentum space.  By the uncertainty principle, this means that the spatial spread $\sigma_x$ of the wavepackets is large compared to the object's de Broglie wavelength $\lambda_{\mathrm{dB}} = \hbar/mv$.

In such a semi-classical approximation, we need only calculate the two possible out-states of the radiative field \emph{conditional} on the two paths.  The evolution takes the form
\begin{align}
\left[ \ket{\psi_1} + \ket{\psi_2} \right] \ket{R_\Omega} \to \ket{\psi_1}\ket{R_1}  + \ket{\psi_2} \ket{R_2}
\end{align}
where $\ket{\psi_i}$ are the (classical) states of the object as it follows path $i = 1,2$, $\ket{R_\Omega}$ is the vacuum  state of the field, $\ket{R_i} = U_i \ket{R_\Omega}$ are the conditional state of the radiation, and $U_i$ is the operator governing the conditional evolution. The reduced density matrix of the object in the two dimensional subspace spanned by $\{ \ket{\psi_i} \}$ is
\begin{align}
\rho = \frac{1}{2} \left( \begin{matrix} 1 & \Gamma \\  \Gamma* & 1 \end{matrix} \right)
\end{align}
where $\Gamma  = \braket{R_1}{R_2}$ is the decoherence factor, given by the overlap of the conditional states of the radiation field.

If the wavelength of the emitted radiation is shorter than the separation between the paths, then the only component of the conditional fields states that overlap will be the vacuum (since any emitted radiation will clearly distinguish one path from the other).  Thus, in the case of $\lambda \lesssim L$, the decoherence factor up to a phase is just the geometric mean of the probabilities that no bremsstrahlung radiation is emitted from the two classical source paths:
\begin{align}
\Gamma &= \left( \sqrt{1-p_1}\bra{\tilde{R}_1} + \sqrt{p_1} \bra{R_\Omega} \right)\left( \sqrt{1-p_2}\ket{\tilde{R}_2} + \sqrt{p_2} \ket{R_\Omega} \right) \\
&= \sqrt{p_1 p_2}.
\end{align}
Here, $\ket{\tilde{R}_i}$ is the component of the conditional states of the field corresponding to non-zero radiation.

Breuer and Pettruccione \cite{Breuer2000, Breuer2001, BreuerBrem} calculate the decoherence due to electromagnetic bremsstrahlung for this experiment and find that interference between the two paths is suppressed by a decoherence factor $\Gamma_{\mathrm{E}}$ where
\begin{align}
\ln \Gamma_{\mathrm{E}} \approx - \frac{2 \alpha_{\mathrm{E}}}{\pi} C \beta^2 
\end{align}
for negligible temperature of the electromagnetic field (i.e.\ initial vacuum state) in the non-relativistic ($\beta \ll 1$) limit.  Above,  $\alpha_{\mathrm{E}} = q^2/4 \pi \hbar c$ is the electromagnetic coupling constant for a clump of charge $q$ and $C$ is a constant of order unity which depends only logarithmically on an infrared cutoff set by the finite duration $\tau$ of the superposition.  In this non-relativistic limit, the photons have wavelength of order $\lambda = \beta^{-1} L \gg L$ and it takes many photons to decohere. For relativistic $\beta$, the wavelength of the radiation is comparable to $L$.  In this case, the decoherence factor $\Gamma$ is the probability that the clump does not emit any photons, which is known to still go like $\ln \Gamma_{\mathrm{E}} \approx - \alpha C' \beta^2$ through basic electromagnetic dipole bremsstrahlung. Notice that, for fixed $L/\tau \sim \beta c$, this does not strongly depend on $L$. 

Gravity has no dipole radiation, unlike electromagnetism, so the primary contributor to bremsstrahlung is quadrupole radiation.  When individual gravitons are sufficient to decohere the superposition, $\lambda \lesssim L$, the decoherence factor $\Gamma$ is just the chance that no gravitons have been emitted.  Calculations of electromagnetic and gravitational quadrupole radiation \cite{Gould1985, Weinberg1965} show that one picks up an extra factor  of $\beta^2$ in the emitted power for all modes, which winds up in the exponent of the probabiity that zero gravitons are emitted.  More formally, one can modify  the calculation of Ford \cite{ford1993electromagnetic, ford1997electromagnetic} (although see Ref.\ \cite{mazzitelli2003decoherence} for corrections) and Breuer and Petruccione \cite{Breuer2001} to show that the factor of $\beta^2$ appears similarly in the non-relativistic case, $\lambda \gg L$.  Essentially, the interaction Hamiltonian is modified as $q u^\mu A_\mu \to \frac{1}{2} m \sqrt{G}  u^\mu u_i^\nu  h_{\mu \nu}$, where $A_\mu$ is electromagnetic, $h_{\mu \nu}$ is  linearized gravity, and $u^\mu$ is the four-velocity of the classical path taken by the object.  The additional copy of $u^\nu$ must appear to contract with the second index in the spin-2 field $h_{\mu \nu}$, and this propagates into a factor of $\beta^2$ in the exponent of the decoherence factor \footnote{An appendix with details will appear in a later version of this paper.}.  The result is
\begin{align}
\label{eq:decoh-exp}
\ln \Gamma_{\mathrm{G}} \approx - \alpha_{\mathrm{G}} C'' \beta^4 = - \frac{G m^2}{\hbar c} C'' \beta^4.
\end{align}
The planck mass 
\begin{align}
m_\mathrm{P} & = \sqrt{\frac{\hbar c}{G}} \\
& \approx 1.2 \times 10^{19} \mathrm{GeV}/c^2 \approx 21 \mu \mathrm{g} \approx 1.3 \times 10^{19}\ \mathrm{amu}
\end{align}
is precisely the mass scale at which the gravitational coupling constant $\alpha_{\mathrm{G}}$ reaches unity and $\Gamma_{\mathrm{G}}$ is driven toward zero.  Thus, the coherent manipulation of Planck-mass superpositions at relativistic speeds will be sensitive to the emission of gravitons through bremsstrahlung emission. Decoherence of non-relativistic objects necessitates an increase in mass of order $\beta^{-2}$. 

Matter interferometry has been demonstrated with $m \sim 10^4$ amu \cite{Gerlich2011,Haslinger2013}, and there are no fundamental obstacles to scaling existing technology to exceed $m \sim10^6$ amu \cite{Haslinger2013, Nimmrichter2011a}.  Since the primary barrier to experiments interfering masses larger than $m \sim 10^7$ amu is the Earth's gravity, spaceborne platforms could push these masses by multiple orders of magnitude (potentially exceeding $m \sim 10^{10}$ amu) using the same fundamental techniques \cite{Nimmrichter2011a, ArndtPC}.  Superpositions of lead spheres with $m \sim10^{14}$ amu \cite{Romero-Isart2011, Romero-Isart2012} and of oscillating mirrors with $m \sim10^{16}$ amu \cite{Marshall2003, Vitali2007} are being pursued, although the spatial extents of such superpositions are too small to decohere through bremsstrahlung.  Given this, the coherent manipulation of Planck-mass ($10^{19}$ amu) objects---though massively difficult---is not inconceivable.  Arguably, it is much more feasible than constructing a detector of Jovian proportions.

The key aspect of this experiment is that the elementary masses making up the lump of matter act together coherently to generate a single graviton.  In order for this to occur, the wavelength must be macroscopic, implying that the momentum carried by the graviton is truly minuscule.  Therefore the emitted graviton is even more difficult to detect by traditional methods than the shorter wavelengths considered elsewhere (but deemed effective invisible \cite{rothman2006can, Boughn2006}).  Nonetheless, its presence can be inferred because decoherence does not require classical influence like momentum or energy transfer.

One might argue that this experiment does not provide evidence for a \emph{particular} graviton emission since the mass formally emits a superposition of graviton number eigenstates, some of which are empty.  Are we not simply measuring that the expected number is larger than zero?

No. Consider a small mass $m \ll m_{\mathrm{P}}$ in the relativistic case where $\lambda \lesssim L$.  This means the coupling to the gravitational field is weak, but that when a single graviton is emitted it in principle provides which-path information (if it could ever be conventionally detected). For simplicity, assume the paths are symmetric so that the largest peak of the interference pattern lies on center of the detector screen.  So long as the magnitude of other peaks does not fall off too quickly, the distance between peaks is determined solely by the de Broglie wavelength of the object.  For sufficiently small $m$ under ideal conditions, the object will never appear in a trough on the detector screen.  But as we increase $m$, gravitons will occasionally be emitted.  When one is, the two paths add incoherently and the object is as likely to found in a trough as at a peak.   If a perfect conventional graviton detector could be built around the experiment, then it would detect an emitted graviton \emph{every time} the object were detected at the bottom of a trough.  Therefore, this technique is no more statistical than traditional single-photon detectors;  it may miss some events but (under idealized conditions) it provides clear evidence for individual graviton emission.  

The non-relativistic case is more subtle, and touches on disagreement over what is necessary for claiming detection of individual quanta.  (See Ref.\ \cite{rothman2006can} for illuminating discussion.)  The photoelectric effect is often taken as clear evidence for the existence of the photon particle, but in fact the relevant experiments actually do not identify individual events.  Rather, one observes a finite frequency threshold necessary for incident light to free electrons from a surface, and then a proportionality between the energy of the freed electrons and the frequency (in excess of the threshold) of the light.  This proportionality constant is of course $\hbar$, and it sets the scale of the electromagnetic quanta even when no individual photons are identified by the experiment.

When a heavy non-relativistic object with $m \beta^2 \gtrsim m_{\mathrm{P}}$ travels through an interference experiment as depicted in Fig.\ \ref{fig:wiggler_diagram}, many gravitons are emitted.  We cannot draw a strong implication from detecting the object in a trough once.  However, collecting many events will still allow us to measure $\hbar$ though \eqref{eq:decoh-exp}, confirming that this is a quantum gravity effect that has no analog in a classical theory of gravity.  (For instance, if a classical gravitational field coupled to the expectation value of the center of mass of the (still quantum) object, no decoherence would result.)  The non-relativistic interferometer experiment is therefore the analog of the photoelectric effect rather than of a single-photon detector.

Of course, such a scheme for detecting the presence of gravitons assumes that all other sources of decoherence can be suppressed. The analysis necessary to confidently attribute observed decoherence to a particular source is discussed elsewhere \cite{Riedel02}, but would doubtlessly be extremely challenging for gravitational Bremmstrahlung. There are at least two irreducible backgrounds that, if they hinder the observation of gravitons, would themselves be exciting new physics: collisional decoherence from relic neutrinos, and bremsstrahlung from fifth forces stronger than gravity.  The investigation of these and other speculative possibilities is deferred to future work.

\bibliographystyle{apsrev4-1}
\bibliography{riedelbib}
\end{document}